# MnAs overlayer on GaN(000$\underline{1}$)-(1x1) – its growth, morphology and electronic structure

B.J. Kowalski, I.A. Kowalik, R.J. Iwanowski, E. Łusakowska, M. Sawicki

*Institute of Physics, Polish Academy of Sciences, Aleja Lotników 32/46,*

*PL-02 668 Warsaw, Poland*

J. Sadowski

*MAX Laboratory, Lund University, PO Box 118, SE-221 00 Lund, Sweden*

*and Institute of Physics, Polish Academy of Sciences, Aleja Lotników 32/46,*

*PL-02 668 Warsaw, Poland*

I. Grzegory, S. Porowski

*High Pressure Research Center, Polish Academy of Sciences,*

*Sokolowska 29, PL-01 141 Warsaw, Poland*

**Abstract**

MnAs layer has been grown by means of MBE on the GaN(000$\underline{1}$)-(1x1) surface. Spontaneous formation of MnAs grains with the diameter of 30-60 nm (as observed by AFM) occurred for the layer thickness bigger than 7 ML. Ferromagnetic properties of the layer with Curie temperature higher than 330 K were detected by SQUID measurements. Electronic structure of the system was investigated *in situ* by resonant photoemission spectroscopy for MnAs layer thickness of 1, 2 and 8 ML. Density of the valence band states of MnAs and its changes due to increase of the layer thickness were revealed.

**Introduction**

Heterostructures consisting of ferromagnetic materials and semiconductors attract much of interest because of their prospective spintronic applications [1]. Among other ferromagnetic systems, selected compounds of Mn (like GaMn and MnAs) are also considered [2-6]. MnAs is especially interesting, not only due to possible applications, but also from a point of view of basic research of magnetic materials. Their magnetic properties are coupled with their structure [7]. This inspired attempts to perform its overgrowth on various substrates (e.g. GaAs(001) [7], GaAs(111) [8]).

In this paper we report a study of MnAs layers deposited by MBE technique on GaN(000$\underline{1}$)-(1x1). Since both materials have hexagonal structure in the plane perpendicular to the *c* axis and $a_{MnAs}>a_{GaN}$, formation of MnAs dots in this system can be expected (contrary to MnAs/GaAs(111) system). The layers of MnAs were deposited stepwise and its electronic structure was investigated *in situ* at each stage of growth (1, 2 and 8 ML) by means of photoelectron spectroscopy. Surface crystallinity was assessed by RHEED and LEED techniques. The morphology of the layer formed by deposition of 8 ML of MnAs was studied *ex situ* by AFM. Its magnetic properties were studied by the SQUID technique.

**Experimental**

The MnAs/GaN system was prepared and its electronic structure was investigated in the National Electron Accelerator Laboratory for Nuclear Physics and Synchrotron Radiation Research MAX-lab, Lund University, Sweden. MnAs layers were grown on bulk GaN substrates by means of MBE technique. The bulk GaN crystals with hexagonal crystalline structure were grown by means of high pressure technique at the High Pressure Research Center, Polish Academy of Sciences, Warsaw, Poland. The (000$\underline{1}$) surface of GaN substrate used in our experiments were initially prepared by *ex situ* mechanical and chemical polishing. Prior to the MnAs growth and photoemission measurements they were introduced to the UHV system and subjected to an *in situ* cleaning procedure consisting of the cycles of Ar$^+$ ion bombardment and subsequent annealing at $500^0$ C. As previously shown [9], such a procedure leads to a clean and well-ordered (1x1) GaN (000$\underline{1}$) surface. That was confirmed for the investigated surfaces by the surface crystallinity assessment done by electron diffraction observations (LEED and RHEED). The MnAs was deposited in the MBE chamber directly attached to the spectrometer. The process was monitored by RHEED. The deposition was realized in three steps. The conditions were fixed to reach subsequently 1, 2, and 8 ML thick overlayer.

The photoemission experiments were carried out at the beamline 41 of MAX-lab. The overall energy resolution was kept around 150 meV, and the angular resolution was about $2^0$. The origin of the binding energy scale was set at the Fermi level (determined for the reference metal sample). The angle between incoming photon beam and the normal to the surface was kept at 45 $^0$. AFM and SQUID studies were performed *ex situ* in the Institute of Physics, Polish Academy of Sciences, Warsaw, Poland.

**Results and discussion**

A stepwise process of MnAs growth enabled us to observe changes in the layer structure as well as in its electronic structure. The first two steps of deposition (1 and 2 ML) caused blurring of the streaked pattern of GaN(000$\underline{1}$)-(1x1). Further deposition led to improvement of the pattern (streaks became again stronger and sharper) - then, at about 7 ML, the critical thickness was achieved and it switched the pattern to a dotted one, indicative of 3D growth. An AFM investigation of the sample surface morphology proved that MnAs grains of the diameter 30-60 nm and the average height of 4 nm were formed (Fig. 1).

Electronic structure of the grown system was investigated *in situ* by means of photoelectron spectroscopy. Since the energy distribution of Mn 3d states is the key issue which must be discussed, resonant photoemission technique was applied in order to resolve Mn 3d contribution to the photoemission spectra against the background of emission from the valence band. Thus, photoelectron experiments were carried out for photon energies close to Mn 3p→3d transition.

When the photon energy fits the energy of the intra-ion transition to the unoccupied states in the partly filled shell (like Mn 3d), two processes leading to the same final state may occur. The first one is the direct photoemission excitation from the open shell to the continuum of free electron states. The second one is a process involving the discrete intra-ion transition (like Mn 3p→3d) followed by a recombination of one of the electrons from the open shell with the hole created in the core level, associated with the emission of another electron from the same shell to the continuum states. For Mn-containing crystals one considers the following transitions:

$$Mn\ 3d^5 + h\nu \rightarrow Mn\ 3d^4 + e^-$$

and

$$Mn\ 3p^6 3d^5 + h\nu \rightarrow Mn\ 3p^5 3d^6 \rightarrow Mn\ 3p^6 3d^4 + e^-.$$

The quantum interference between these processes manifests itself in a resonant increase of the photoemission from the open shell. Thus, resonant photoemission is widely used for identification of the features of the photoemission spectra which can be ascribed to the emission from partly filled shells.

Fig. 2a shows selected photoemission spectra acquired at photon energies of 50 and 51 eV (i.e. under resonance conditions). They were measured for clean GaN(000$\underline{1}$)-(1x1) surface and for the same surface with 8 ML thick layer of MnAs. Analysis of the full sets of spectra, taken for photon energies from 45 to 65 eV, showed that the resonance photoemission conditions occurred for 50 and 51 eV. For these energies enhancement in the whole valence band was observed. However, two features, at 4.1 and 6.7 eV, became particularly strong for 51 and 50 eV, respectively (Fig. 2a). This suggests that Mn in the system occurs in two different states with different resonance energies. It is well known from studies of various Mn compounds [10,11] that resonance energy as well as the shape of Mn 3d related emission strongly depend on chemical identity of Mn-atom ligands and symmetry of their arrangement. The spectrum obtained for MnAs(8ML)/GaN at hν = 51 eV is very similar to that taken for 1 ML thick MnAs layer at the same photon energy. Moreover, for the thin layer (1 or 2 ML) the spectra at hν = 50 and 51 eV are much alike. The feature at the binding energy of 6.7 eV becomes discernible only for the MnAs layer with the thickness of 8 ML (at hν = 50 eV). Thus, Mn atoms exhibiting the resonance energy close to 51 eV seem to be built into the region adjacent to the MnAs/GaN interface. Those with 50 eV resonance energy are probably located in MnAs grains. Spin-resolved photoemission studies of MnAs epitaxial layers grown on GaAs [12] revealed Mn 3d majority spin related contribution with only a maximum at 3.7 and a shoulder at 1.9 eV. The feature at 6.7 eV appearing in our spectra together with the change of the sample morphology seems to be related to substantial change in the Mn atom neighbourhood.

For the spectra of MnAs/GaN, the relatively strong Ga 3d peak at about 20 eV shows that we still record some emission from the substrate (see Fig. 2a). Thus, one had to subtract the spectrum of GaN from that of MnAs/GaN in order to reveal emission from the MnAs layer. The curves were normalized with respect to the Ga 3d peak. Fig. 2b shows a difference spectrum obtained for hν = 50 eV. This curve reproduces the shape of the electronic states distribution of MnAs, with enhanced Mn 3d contribution due to resonance conditions of the measurement – this refers especially to the features at 4.1 and 6.7 eV.

Interaction between deposited species and GaN were also monitored by spectroscopy of selected core levels. Analysis of photoemission spectra of the Ga 3d core level (Fig. 3) revealed no additional features which could be indicative of substantial GaN surface disruption during deposition.

**Conclusions**

MnAs layer grown by means of MBE on the GaN(000$\underline{1}$)-(1x1) surface spontaneously formed grains with the diameter of 30-60 nm, for the layer thickness bigger than 7 ML. Ferromagnetic properties were detected in the system by SQUID measurements. However, the magnetic moment vs. temperature dependence is indicative for $T_C$ much higher than that of NiAs-type MnAs (313 K). Thus, we have to admit that substantial contribution of other Mn compounds at the interface (like MnGa) dominates the observed magnetic moment of the sample, although the Ga 3d spectroscopy showed that amount of MnGa should be very small.

The shape of the electronic states distribution in MnAs was determined by resonant photoemission spectroscopy at the Mn 3p→3d transition. Presence of manganese atoms with two different ligand configurations in the MnAs layer was revealed.

**Acknowledgments** This work was supported by KBN (Poland) project 2P03B 046 19 and by the European Commission programs ICAI-CT-2000-70018 (Center of Excellence CELDIS) and G1MA-CT-2002-4017 (Center of Excellence CEPHEUS) and G5RD-CT-2001-0535 (FENIKS).

**Figure captions:**

Fig.1. Surface morphology of 8 ML MnAs grown on GaN(000$\underline{1}$)-(1x1), obtained by AFM.

Fig. 2. Photoemission spectra taken at photon energies of 50 and 51 eV for MnAs(8ML)/GaN compared with those of clean GaN surface (a). Difference spectrum corresponding to the emission from the MnAs overlayer (b).

Fig. 3. The Ga 3d core level peak measured for clean GaN and MnAs(8ML)/GaN.

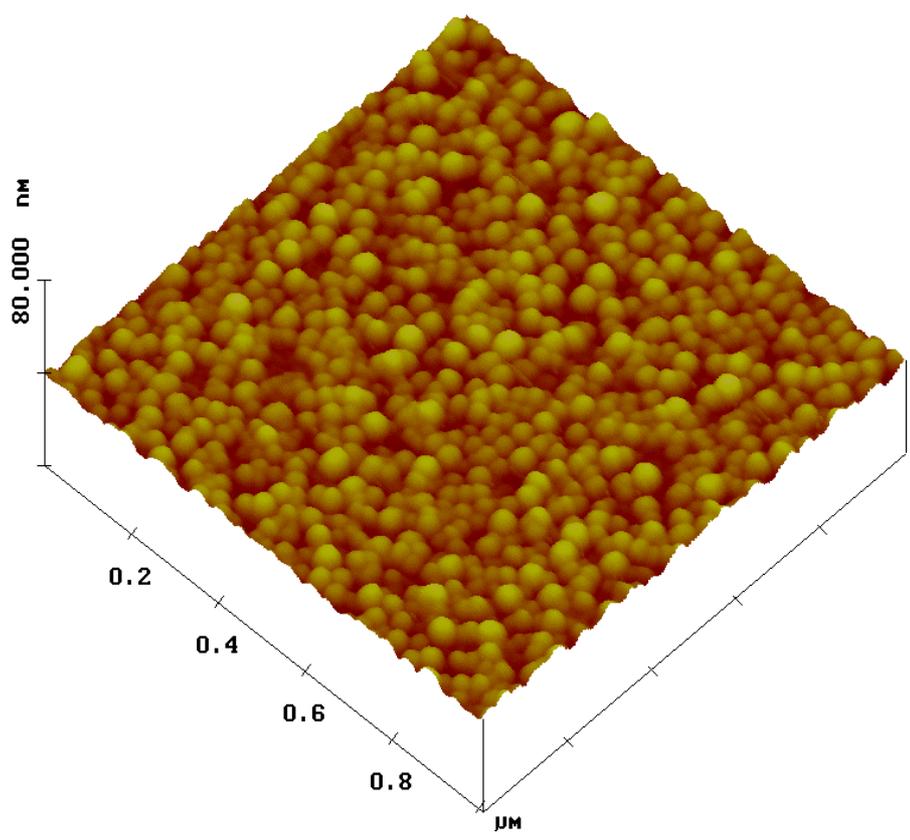

a.

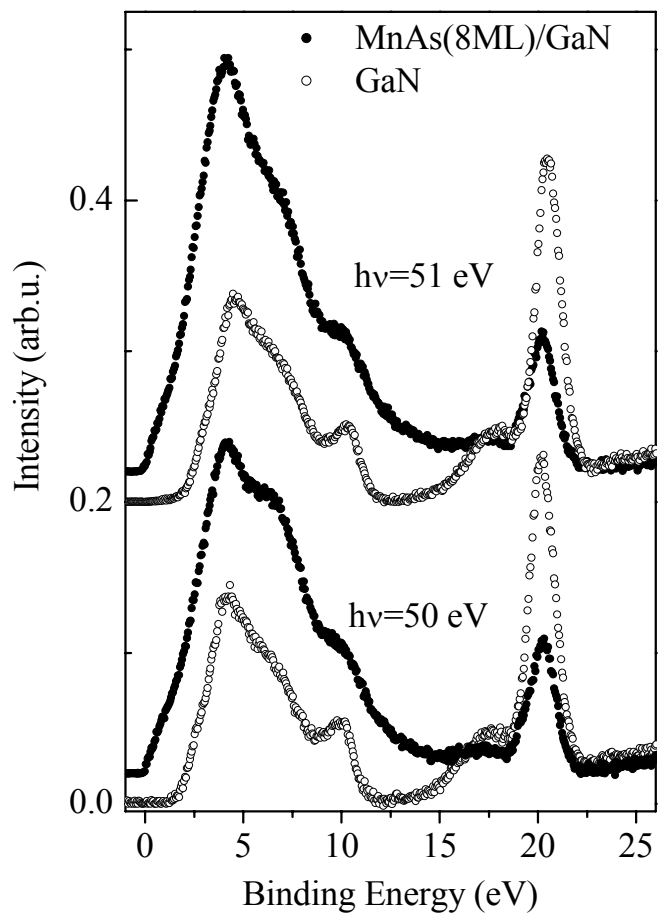

b.

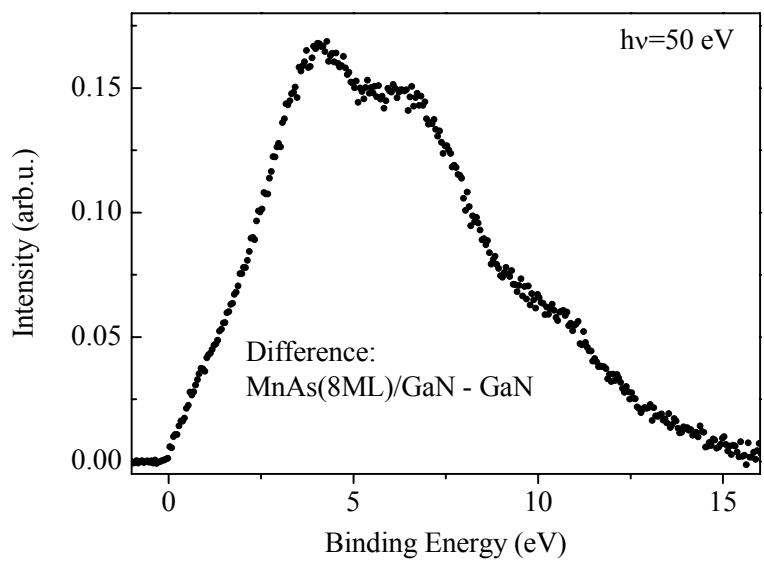

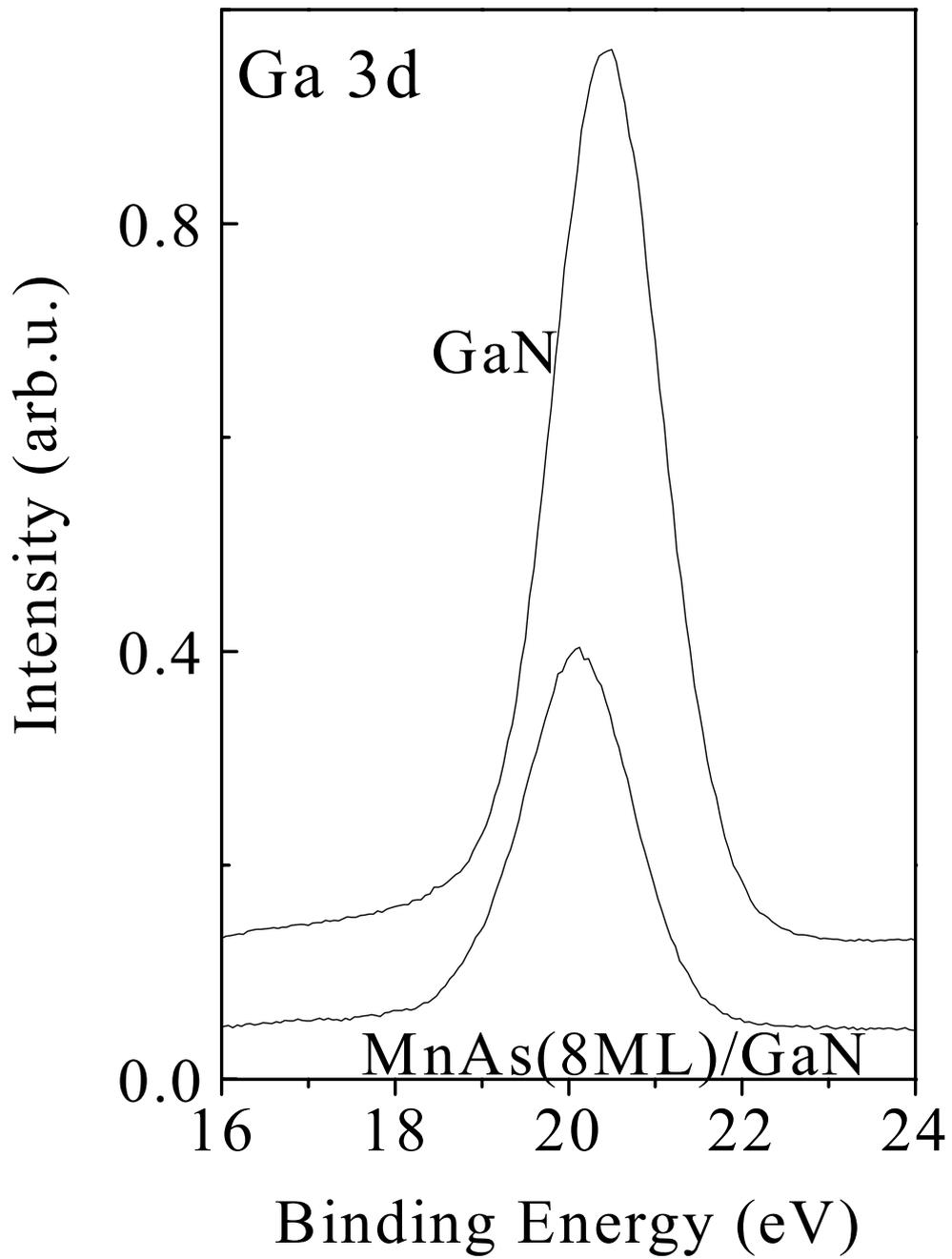